\documentclass[journal,twoside,web]{ieeecolor}
\usepackage{generic}
\usepackage{cite}
\usepackage{amsmath,amssymb,amsfonts}
\usepackage{algorithmic}
\usepackage{graphicx}
\usepackage{algorithm,algorithmic}
\usepackage{hyperref}
\hypersetup{hidelinks=true}
\usepackage{textcomp}

\usepackage[utf8]{inputenc}         
\usepackage[english]{babel}
\usepackage{multirow}
\usepackage{multicol}
\usepackage{soul}
\usepackage[normalem]{ulem}
\usepackage{xcolor,colortbl}
\usepackage{epstopdf}
\usepackage{cite}
\usepackage{capt-of}

\usepackage{tabularx}
\usepackage{bm}
\usepackage{calc}

\usepackage{xcolor,colortbl}
\usepackage{tabu}
\usepackage{mdframed}
\usepackage{url}
\usepackage{pifont}
\usepackage{balance}

\usepackage{array}
\usepackage{rotating}

\usepackage{makecell}
\usepackage{subcaption}
\usepackage{textcomp}

\def\BibTeX{{\rm B\kern-.05em{\sc i\kern-.025em b}\kern-.08em
		T\kern-.1667em\lower.7ex\hbox{E}\kern-.125emX}}
\begin{document}
	\title{L-SeqSleepNet: Whole-cycle Long Sequence Modelling for Automatic Sleep Staging}
	\author{{Huy~Phan,
			Kristian P. Lorenzen,
			Elisabeth Heremans, 
			Oliver~Y.~Ch\'{e}n, 
			Minh C. Tran, \\
			Philipp Koch,
			Alfred Mertins,
			Mathias Baumert,
			Kaare B. Mikkelsen,
			and~Maarten~De~Vos
			\thanks{H. Phan is with Amazon Alexa, Cambridge, MA 02142, United States. K. P. Lorenzen and K. Mikkelsen are with the Department of Electrical and Computer Engineering, Aarhus University, Aarhus 8200, Denmark. E. Heremans and M. De Vos are with the Department of Electrical Engineering and the Department of Development and Regeneration, KU Leuven, 3001 Leuven, Belgium. O. Y. Ch\'{e}n is with the Department of Laboratory Medicine and Pathology (DMLP), Lausanne University Hospital (CHUV), Lausanne, Switzerland and the Faculty of Biology and Medicine (FBM), University of Lausanne, Lausanne, Switzerland. M. C. Tran is with Nuffield Department of Clinical Neurosciences, University of Oxford, Oxford OX3 9DU, UK. P. Koch and A. Mertins are with the Institute for Signal Processing, University of L\"ubeck, L\"ubeck 23562, Germany and the German Research Center for Artificial Intelligence (DFKI), L\"ubeck 23562, Germany. M. Baumert is with School of Electrical and Electronic Engineering, The University of Adelaide, Adelaide SA 5005, Australia.}
			\thanks{The work was done when H. Phan was at the School of Electronic Engineering and Computer Science, Queen Mary University of London, UK and the Alan Turing Institute, UK and prior to joining Amazon.}
			\thanks{Source code is available at \href{http://github.com/pquochuy/l-seqsleepnet}{http://github.com/pquochuy/l-seqsleepnet}.}
			\thanks{Corresponding author: {\tt\footnotesize huypq@amazon.co.uk}}
	}}
	
	\maketitle
	
	\begin{abstract}
		Human sleep is cyclical with a period of approximately 90 minutes, implying long temporal dependency in the sleep data. Yet, exploring this long-term dependency when developing sleep staging models has remained untouched. In this work, we show that while encoding the logic of a whole sleep cycle is crucial to improve sleep staging performance, the sequential modelling approach in existing state-of-the-art deep learning models are inefficient for that purpose. We thus introduce a method for efficient long sequence modelling and propose a new deep learning model, L-SeqSleepNet, which takes into account whole-cycle sleep information for sleep staging. Evaluating L-SeqSleepNet on four distinct databases of various sizes, we  demonstrate state-of-the-art performance obtained by the model over three different EEG setups, including scalp EEG in conventional Polysomnography (PSG), in-ear EEG, and around-the-ear EEG (cEEGrid), even with a single EEG channel input. Our analyses also show that L-SeqSleepNet is able to alleviate the predominance of N2 sleep (the major class in terms of classification) to bring down errors in other sleep stages. Moreover the network becomes much more robust, meaning that for all subjects where the baseline method had exceptionally poor performance, their performance are improved significantly.  Finally, the computation time only grows at a sub-linear rate when the sequence length increases. 
	\end{abstract}
	
	
	\section{Introduction}
	\label{sec:introduction}
	
	Sleep is a slow-transitioning neural process, and thus, the data recorded from this process embeds abundant sequential information. Capturing this sequential information has been shown to be crucial for automatic sleep staging systems to achieve good performance. In fact, the capacity of sequential modelling has been the driving force behind existing deep-learning-based sleep staging models, bringing the machine scoring performance on par with that of human experts \cite{Phan2022b}. Using recurrent neural networks (RNNs) (e.g., Long Short-Term Memory (LSTM) \cite{Hochreiter1997}) or, more recently, the Transformer architecture \cite{Vaswani2017}, these models are able to capture the temporal dependency in a sequence of multiple consecutive epochs of sleep data, resembling the way a human expert conducts manual scoring. The sequence length (i.e., the number of epochs in the sequence) was shown to be important. Many different works \cite{Phan2019a, Supratak2017, Seo2020, Phan2022c, MousaviI2019} found a length around 20-30 epochs to be effective and these have been \emph{de facto} values for this hyper-parameter. At least, using a longer sequence was reported to lead to little to no performance gain at the cost of significantly increased computational overhead \cite{Phan2019a, Supratak2017}.
	
	\setlength\tabcolsep{2.25pt}
	\begin{table}[!t]
		\caption{Overall accuracy of SeqSleepNet and SleepTransformer obtained on the SHSS database with a long sequence length of \{100, 200\} compared to suggested values, 20 (SeqSleepNet) and 21 (SleepTransformer). $^*$Note that this result is slightly different from that reported for SeqSleepNet in \cite{Phan2022a} as we did not exercise early stopping here.}
		\begin{center}
			\begin{tabular}{|>{\arraybackslash}m{1.2in}|>{\centering\arraybackslash}m{0.35in}|>{\centering\arraybackslash}m{0.6in}|>{\centering\arraybackslash}m{0.6in}|>{\centering\arraybackslash}m{0in} @{}m{0pt}@{}}
				\cline{1-4}
				\multirow{2}{*}{} & \multicolumn{3}{c|}{Sequence length} &  \parbox{0pt}{\rule{0pt}{1ex+\baselineskip}} \\ [0ex]  	
				\cline{2-4}
				& 20/21 & 100 & 200  & \parbox{0pt}{\rule{0pt}{0.5ex+\baselineskip}} \\ [0ex]  	
				\cline{1-4}
				SeqSleepNet  \cite{Phan2019a} & $87.2^*$ & $87.4~\textcolor{blue}{(\uparrow 0.2)}$  & $87.5~\textcolor{blue}{(\uparrow 0.3)}$ & \parbox{0pt}{\rule{0pt}{0ex+\baselineskip}} \\ [0ex]  	
				SleepTransformer \cite{Phan2022c}  & $87.7$& $86.6~\textcolor{red}{(\downarrow 1.1)}$  & $86.3~\textcolor{red}{(\downarrow 1.4)}$ & \parbox{0pt}{\rule{0pt}{0ex+\baselineskip}} \\ [0ex]  	
				\cline{1-4}
			\end{tabular}
		\end{center}
		\label{tab:long_seq}
	\end{table}
	
	We conducted some initial experiments with a large sequence length of \{100, 200\} to verify the above observation. For this investigation, we employed two models, SeqSleepNet \cite{Phan2019a} and SleepTransformer \cite{Phan2022c}, and the SHHS database \cite{Zhang2018, Quan1997}, a large database consisting of recordings from 5,791 subjects. Both models conform to the state-of-the-art sequence-to-sequence sleep staging framework \cite{Phan2022b} but the former uses LSTM and the latter uses Transformer for sequential modelling purposes. The obtained overall staging accuracy in Table \ref{tab:long_seq} indeed attest the negligible possible impact of a long sequence length in case of SeqSleepNet, marginally improving the accuracy by $0.2-0.3\%$ when the sequence length is 5 and 10 times longer than the typical value (i.e., 20 epochs). Even worse, an adverse effect is observed in the case of SleepTransformer whose accuracy noticeably drops $>\!1.0\%$ when the sequence length increases from 21 to 100 and 200 epochs. This is likely because the Transformer-based architecture usually needs a lot more data to train, and thus, is more prone to overfitting when the receptive field gets larger with the increased sequence length. These results  consolidate what was observed in previous works \cite{Phan2019a, Supratak2017}.
	
	However, in the particular context of sleep data, such a negative effect of a long sequence length to the automatic staging performance appears to be implausible. Typically, a person goes through four to six sleep cycles per night. One complete sleep cycle takes roughly 90 to 110 minutes (equivalent to 180 to 220 epochs of 30 seconds each), transitioning through Awake$\rightarrow$N1$\rightarrow$N2$\rightarrow$N3$\rightarrow$REM sequentially and the time spent in each stage is well-studied \cite{Patel2022}. Furthermore, there are temporal dynamic processes that underpin the sleep cycle. Several phenomena can be observed from sleep's physiological signals, reflecting these dynamic processes. N2 sleep, for example, at the beginning of a cycle is not the same as at the end of the cycle \cite{Purcell2017}. When looking at the cyclic alternating patterns (CAP), for instance, there are more A phases before REM sleep onset than after a REM period \cite{Hartmann2019, Hartmann2020}. Also, the cycles themselves differ with more REM in the morning.	All in all, sleep cycles exhibit temporal sleep-transitioning structures specific to the sleep process. The implication of this is that the temporal interdependence in sleep data could be as long as a whole cycle. For instance, intuitively, knowing that an epoch is N1 should increase the likelihood of an epoch 25 minutes after it to be N2, given the fact that N1 lasts between 1-5 minutes and N2 lasts $\ge25$ minutes.
	
	We argue that the negative effects of long sequences observed so far in the literature are due to model deficiency. That is, considering a long sequence as a ``flat'' sequence \emph{per se}, the typical sequential modelling method in existing models \cite{Phan2022b} is incapable of handling long sequences, e.g., up to one whole sleep cycle. We hypothesize that an appropriate method for equipping a sleep staging system with long sequential modelling capability would benefit its performance. The present work introduces a new method that is able to model long sequences (i.e. one sleep cycle or more) to achieve new state-of-the-art performance, even with a single-EEG input.

	\section{Materials}
	\label{sec:materials}
	
\begin{figure} [!t]
		\centering
		\includegraphics[width=1\linewidth]{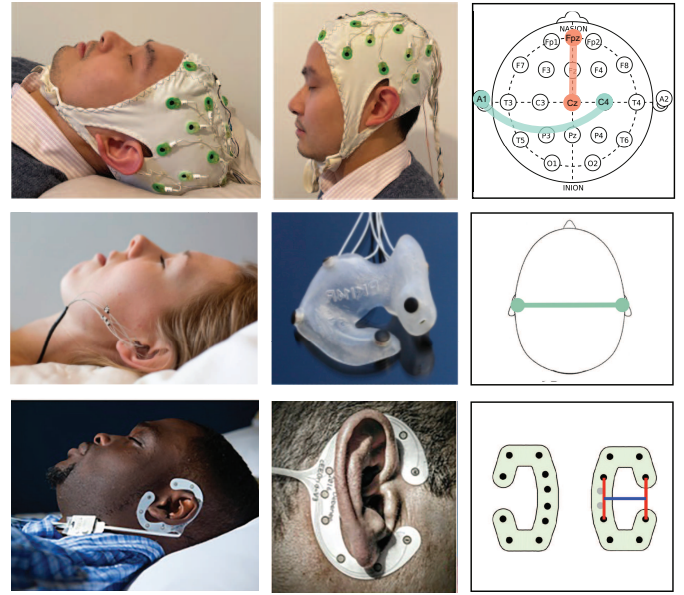}
		\caption{Illustration of the three EEG setups: scalp EEG (top row), in-ear EEG (ear-EEG) (middle row), and around-the-ear EEG (cEEGrid) (bottom row).}
		\label{fig:eeg_setups}
	\end{figure}

	We employed four databases in this work. On the one hand, the SHHS and SleepEDF databases are based on conventional PSG setup from which scalp EEG derivations, C4-A1 for SHHS and Fpz-Cz for SleepEDF, were derived (see Figure~\ref{fig:eeg_setups}, top row). On the other hand, the ear-EEG and cEEGrid databases are based on in-ear EEG setup \cite{Mikkelsen2019b, Mikkelsen2021} (see Figure~\ref{fig:eeg_setups}, middle row) and around-the-ear EEG setup \cite{Debener2015,Debener2012} (see Figure \ref{fig:eeg_setups}, bottom row), respectively. 
	
	{\bf SHHS:} This large database was gathered from multiple centers  as part of the clinical trial ``Sleep Heart Health Study (SHHS)'', ClinicalTrials.gov number NCT00005275 to study the effect of sleep-disordered breathing on cardiovascular diseases \cite{Zhang2018, Quan1997}. It consists of two sets of PSG recordings, namely Visit 1 and Visit 2. Here, we employed Visit 1 consisting of 5,791 PSG recordings from 5,791 subjects, aged 39-90. Following \cite{Sors2018}, we excluded those recordings without the presence of all five sleep stages. As a result, 5,463 PSG recordings were retained. In addition, we also experiment the proposed network on all 5,791 subjects without data exclusion to lay the ground for future comparison. The recordings were manually scored following the R\&K guidelines \cite{Hobson1969} where each 30-second epoch was labelled as one of eight categories \{W, N1, N2, N3, N4, REM, MOVEMENT, UNKNOWN\}. In our experiments, N3 and N4 stages were merged and considered as N3 collectively. MOVEMENT and UNKNOWN epochs were discarded. We adopted C4-A1 EEG in the experiments.
	
	{\bf SleepEDF:} This is the Sleep Cassette subset of the Sleep-EDF Expanded dataset \cite{Kemp2000, Goldberger2000} (version 2013). It  consists of 20 subjects (10 males and 10 females) aged 25-34. Each subject had two consecutive day-night PSG recordings recorded, except for the subject 13 whose one night's data was lost due to device failure, making a total of 39 PSG recordings. This database was manually labelled according to the R\&K guideline \cite{Hobson1969} where each 30-second epoch was labelled as one of eight categories \{W, N1, N2, N3, N4, REM, MOVEMENT, UNKNOWN\}. Similar to SHHS, N3 and N4 stages were merged and considered as N3 collectively while MOVEMENT and UNKNOWN categories were excluded. We adopted the Fpz-Cz EEG channel in the experiments. Adhering to the common setting in literature, a recording was trimmed starting from 30 minutes before to 30 minutes after its \emph{in-bed} part.
	
	{\bf ear-EEG:} This database constitutes ear-EEG recordings of 20 subjects (7 males and 13 females, aged 22-36) recorded using the same ear-EEG equipment. In relation to sleep apnea, all subjects were scored 1 or less after answering to the STOP-BAND questionnaire \cite{Hanquist2018}. Each subject had four nights of recordings. Three recordings were excluded after artefact rejection \cite{Mikkelsen2021} and two other recordings were excluded as their remaining lengths was less than 200 epochs after artefact rejection. This resulted in 75 recordings in total. The labels of the data were obtained via manual scoring of the PSG recordings which were recorded concurrently to the ear-EEG as a reference. Manual scoring was done by two independent and experienced sleep technicians according to the AASM guidelines \cite{berry_aasm_2016} where each 30-second epoch is labelled as one in five categories \{Wake, N1, N2, N3, REM\}. As in \cite{Mikkelsen2021}, we used the labels from the scorer 1 as the ground truth here. We adopted the bilateral ear-EEG derivation (i.e. the average of the left ear electrodes relative to the average of the right ear electrodes (see Figure \ref{fig:eeg_setups}, middle row, right picture)) in the experiments. More details about the recording setup and data preprocessing can be found in \cite{Mikkelsen2019b, Mikkelsen2021}. 
	
	{\bf cEEGrid:} This database \cite{Mikkelsen2019, Sterr2018} was recorded at the University of Surrey using a lightweight flex–printed electrode strip, namely the cEEGrid array \cite{Debener2015,Debener2012}, fitted behind the ear, as illustrated in Figure \ref{fig:eeg_setups} (bottom row, left and middle pictures). 20 subjects, 8 males and 12 females aged 34.9$\pm$13.8 years, took part in the data recording and one overnight cEEGrid recording was recorded for each subject. Two recordings were lost due to human error and six recordings were excluded because of excessive artefacts and data missing. 12 remaining recordings were retained and used in the experiments as in \cite{Phan2020b}. They were mostly good sleepers \cite{Sterr2018}. The labels of the data were obtained via manual scoring of the PSG recordings which were recorded concurrently as reference for the cEEGrid data \cite{Sterr2018}. The FB(R) (``front versus back'') derivation for the right ear (see Figure \ref{fig:eeg_setups}, bottom row, right picture) which was the best derivation \cite{Mikkelsen2019}, was adopted for the experiments. More details about the recording setup and data preprocessing can be found in \cite{Mikkelsen2019, Sterr2018}.
	
	\begin{figure*} [!t]
		\centering
		\includegraphics[width=0.9\linewidth]{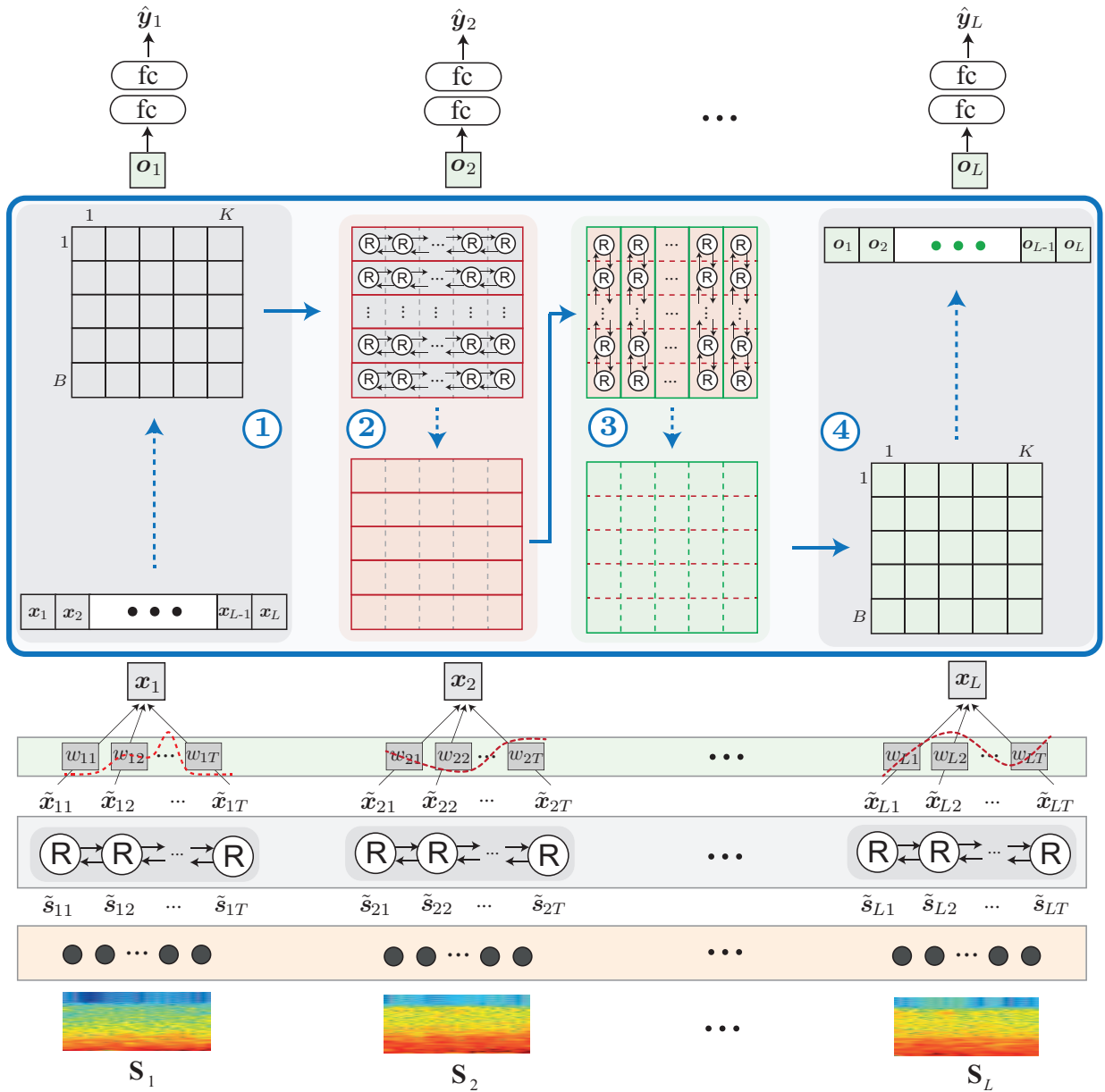}
		\caption{The architecture of L-SeqSleepNet. The long sequence modelling happens in the big blue box, consisting of 4 steps: \textcircled{1} the long sequence of length $L$ is folded into B non-overlapping subsequences of length $K$; \textcircled{2} sequential modelling within subsequences, reflected by the change of the grid's color from grey to red; \textcircled{3} sequential modelling across subsequences, reflected by the change of the grid's color from red to green; \textcircled{4} the subsequences are unfolded to resume the original sequence of length $L$.}
		\label{fig:lseqsleepnet}
	\end{figure*}
	
	\section{L-SeqSleepNet with Long Sequence Modelling Capability}
	
	Let $\{\mathcal{S}_n\}_{n=1}^N$ denote a training set of size $N$ where $\mathcal{S}_n = \left((\bm{S}^{(n)}_1, \bm{y}^{(n)}_1), \ldots, (\bm{S}^{(n)}_L, \bm{y}^{(n)}_L)\right)$ is the $n$-th sequence consisting of $L$ consecutive sleep epochs. $\bm{S}^{(n)}_\ell$ and $\bm{y}^{(n)}_\ell\!\in\!\{0,1\}^C$ represent the $\ell$-th 30-second sleep epoch and its one-hot encoding label in the $n$-th sequence, respectively. Here, $C=5$ as we are dealing with 5-stage sleep staging. Similar to a sequence-to-sequence sleep staging model \cite{Phan2022b}, given a sequence $(\bm{S}_1, \bm{S}_2, \ldots, \bm{S}_L)$ as input, L-SeqSleepNet aims to classify all the epochs in the input sequence at once and produce the  sequence of probability output vectors $(\bm{\hat{y}}_1, \ldots ,\bm{\hat{y}}_L)$, where $\bm{\hat{y}}^{(n)}_\ell\!\in\![0,1]^C$, $1\le\!\ell\!\le\!L$, is for the $\ell$-th epoch. However, different from existing sequence-to-sequence sleep staging models that consider short sequences ($L$ between 20-30 epochs or 10-15 minutes equivalently), we are interested in long sequences (e.g. $L\!=\!200$ or 100 minutes equivalently) so that a sequence roughly covers an entire sleep cycle.
	
	The architecture of L-SeqSleepNet is illustrated in Figure \ref{fig:lseqsleepnet}. It receives the time-frequency input and has the epoch encoding part inherited from SeqSleepNet \cite{Phan2019a} while the sequence encoding part is devised to handle long sequences efficiently. For completeness, we describe all of these components in order in the following sections.
	
	\subsection{Input} 
	
	The EEG signal of a 30-second epoch is converted into a log-magnitude time-frequency image $\bm{S}$ with $T\!=\!29$ time steps and $F\!=\!129$ frequency bins.  To that end, short-time Fourier transform (STFT) is applied to the signal with a window length of 2 seconds and 50\% overlap. In addition, Hamming window and 256-point fast Fourier transform (FFT) are used. The obtained amplitude spectrum is then log-transformed to result in the image $\bm{S} \in \mathbb{R}^{T\times F}$.
	
	\subsection{Epoch encoding}
	\label{ssec:epoch_encoding}
	The role of the epoch encoding component is to learn the feature map, $\mathcal{F}(\bm{S}): \bm{S} \mapsto \bm{x}$, in order to transform an input epoch $\bm{S}$ into a high-level feature vector $\bm{x}$ for representation. This is realized by a subnetwork which is shared across all epochs in an input sequence. The subnetwork is composed of (i) a learnable filterbank layer, (ii) a bidirectional Long Short-Term Memory (BLSTM) \cite{Hochreiter1997}, and (iii) a gated attention layer. The \emph{learnable} filterbank layer \cite{Phan2018a} consists of $M$ filters ($M < F$), being tasked to smooth and reduce the frequency dimension from $F$ to $M$ bins. The resulting image $\bm{\tilde{S}}$ of size $T\times M$ is then treated as a sequence of $T$ vectors (i.e., $T$ image columns), $(\bm{\tilde{s}}_1, \bm{\tilde{s}}_2, \ldots, \bm{\tilde{s}}_T)$, where $\bm{\tilde{s}}_t \in \mathbb{R}^M$, $1 \le t \le T$. In order to capture the sequential information at the epoch level, this sequence is encoded by the BLSTM with recurrent batch normalization \cite{Cooijmans2016}, into a sequence of vectors $(\bm{\tilde{x}}_1, \bm{\tilde{x}}_2, \ldots, \bm{\tilde{x}}_T)$:
	\begin{align}
	(\bm{\tilde{x}}_1, \bm{\tilde{x}}_2, \ldots, \bm{\tilde{x}}_T) = BLSTM_e(\bm{\tilde{s}}_1, \bm{\tilde{s}}_2, \ldots, \bm{\tilde{s}}_T).
	\end{align}
	Here, $\bm{\tilde{x}}_t \in \mathbb{R}^{H_e}$ with $\frac{H_e}{2}$ is the size of the hidden states in $BLSTM_{e}$. We use the subscript $e$ to indicate modelling at the \emph{epoch} level and distinguish it from other BLSTMs in Section \ref{ssec:long_seq_model}. Afterwards, the gated attention layer \cite{Luong2015b} is learned to produce attention weights $(w_1,w_2,\ldots,w_T)$ which are used to combine the feature vectors $(\bm{\tilde{x}}_1, \bm{\tilde{x}}_2, \ldots, \bm{\tilde{x}}_T)$ to derive the embedding vector $\bm{x} \in \mathbb{R}^{H_e}$ representing the input epoch $\bm{S}$:
	\begin{align}
	\bm{x} = \sum_{t=1}^Tw_t\bm{\tilde{x}}_t,
	\end{align}
	where
	\begin{align}
	w_t = \frac{\exp(\bm{u}_t^\mathsf{T}\bm{a})}{\sum_{i=1}^T\exp(\bm{u}_i^\mathsf{T}\bm{a})}, \\
	\bm{u}_t = \tanh(\bm{W}_a\bm{\tilde{x}}_t + \bm{b}_a).
	\end{align}
	In above equations, $\bm{W}_a \in \mathbb{R}^{A\times H_e}$ and $\bm{b}_a \in \mathbb{R}^A$ are trainable weight matrix and bias vector, respectively. $\bm{a} \in \mathbb{R}^A$ is the trainable context vector and $A$ is the so-called attention size. After the epoch encoding subnetwork described in Section \ref{ssec:epoch_encoding}, the input sequence $(\bm{S}_1, \bm{S}_2, \ldots, \bm{S}_L)$ has been transformed into the sequence of embeddings $(\bm{x}_1, \bm{x}_2, \ldots, \bm{x}_L)$. 
	
	\subsection{Long sequence modelling}
	\label{ssec:long_seq_model}
	Encoding the sequential information in the sequence of epoch-wise feature vectors $(\bm{x}_1, \bm{x}_2, \ldots, \bm{x}_L)$ has proved to be the key behind the success of existing sequence-to-sequence sleep staging models \cite{Phan2022b}. This has been commonly accomplished by a subnetwork with sequential modelling capacity, such as RNN \cite{Supratak2017, Phan2019a, Guillot2021, Seo2020, Olesen2021} or Transformer \cite{Phan2022c}. However, we have shown earlier that this approach is inefficient to handle long sequences. 
	
	Central to L-SeqSleepNet's architecture is the subnetwork (the big blue box in Figure \ref{fig:lseqsleepnet}) that is capable of long sequence modelling. In intuition, the processing of this component is composed of four steps indicated by the circled numbers in the figure: \emph{folding}, \emph{intra-subsequence sequential modelling}, \emph{inter-subsequence sequential modelling}, and \emph{unfolding}. We firstly fold the long sequence of length $L$ into $B$ non-overlapping subsequences of length $K$, where $L = B \times K$. Sequential modelling is then performed within each of the subsequences (i.e. intra-subsequence sequential modelling), followed by sequential modelling across the subsequences (i.e. inter-subsequence sequential modelling). Eventually, the subquences are unfolded to resume the long sequence of original length $L$.
	
	Formally, assume that we have folded the sequence $(\bm{x}_1, \bm{x}_2, \ldots, \bm{x}_L)$ into $B$ non-overlapping subsequences of size $K$:
	\begin{align}
	\left( \begin{array}{cccc} \bm{{x}}^{(1)}_1 & \bm{{x}}^{(1)}_2 & \ldots & \bm{{x}}^{(1)}_K \\
	\bm{{x}}^{(2)}_1 & \bm{{x}}^{(2)}_2 & \ldots & \bm{{x}}^{(2)}_K \\
	\ldots & \ldots & \ldots & \ldots \\
	\bm{{x}}^{(B)}_1 & \bm{{x}}^{(B)}_2 & \ldots & \bm{{x}}^{(B)}_K \end{array} \right).
	\end{align}
	Here, we use the subscript $k$, $1 \le k \le K$, to indicate the index of an element inside a subsequence and the superscript $b$, $1 \le b \le B$, to indicate the index of a subsequence. After folding, the $\ell$-th element in the original sequence will become the $k$-th element in the $b$-th subsequence, where 
	\begin{align}
	b &= \left\lfloor\frac{\ell - 1}{K}\right\rfloor + 1,  \\
	k &= \left[(\ell-1)\!\!\mod K\right] + 1.
	\end{align}
	
	Intra-subsequence sequential modelling (along horizontal direction as illustrated in Figure \ref{fig:lseqsleepnet}) is carried out on a $b$-th subsequence $(\bm{x}^{(b)}_1, \bm{x}^{(b)}_2, \ldots, \bm{x}^{(b)}_K)$ using a BLSTM with recurrent batch normalization, transforming it into a subsequence of output vectors $(\bm{\tilde{o}}^{(b)}_1, \bm{\tilde{o}}^{(b)}_2, \ldots, \bm{\tilde{o}}^{(b)}_K)$:	 
	\begin{align}
	(\bm{\tilde{o}}^{(b)}_1, \bm{\tilde{o}}^{(b)}_2, \ldots, \bm{\tilde{o}}^{(b)}_K) = BLSTM_{ss}(\bm{x}^{(b)}_1, \bm{x}^{(b)}_2, \ldots, \bm{x}^{(b)}_K),
	\end{align}
	where $\bm{\tilde{o}}^{(b)}_k \in \mathbb{R}^{H_{ss}}$. $\frac{H_{ss}}{2}$ is the size of the hidden states in $BLSTM_{ss}$. The subscript $ss$ is used to indicate the modelling at the \emph{subsequence} level. The output vectors $\bm{\tilde{o}}^{(b)}_k$ are then linear transformed via a fully connected (fc) layer, followed by layer normalization (LN) \cite{Ba2016} and a residual connection:
	\begin{align}
	\bm{\bar{o}}^{(b)}_k = \bm{\tilde{o}}^{(b)}_k + LN(\bm{W}_{ss}\bm{\tilde{o}}^{(b)}_k + \bm{b}_{ss}).
	\end{align}
	Here, $\bm{W}_{ss} \in \mathbb{R}^{H_{ss} \times H_{ss}}$ and $\bm{b}_{ss} \in \mathbb{R}^{H_{ss}}$ denote the trainable weight matrix and bias vector of the fc layer, respectively. As a result, we obtain the following $B$ output subsequences:
	\begin{align}
	\left( \begin{array}{cccc} \bm{\bar{o}}^{(1)}_1 & \bm{\bar{o}}^{(1)}_2 & \ldots & \bm{\bar{o}}^{(1)}_K \\
	\bm{\bar{o}}^{(2)}_1 & \bm{\bar{o}}^{(2)}_2 & \ldots & \bm{\bar{o}}^{(2)}_K \\
	\ldots & \ldots & \ldots & \ldots \\
	\bm{\bar{o}}^{(B)}_1 & \bm{\bar{o}}^{(B)}_2 & \ldots & \bm{\bar{o}}^{(B)}_K \end{array} \right).
	\end{align}
	Up to this point, each output vector $\bm{\bar{o}}^{(b)}_k\!\!\in\!\mathbb{R}^{H_{ss}}$ in a $b$-th subsequence is expected to contain the information of the entire subsequence.
	
	Inter-subsequence sequential modelling (along vertical direction as illustrated in Figure \ref{fig:lseqsleepnet}) is then conducted at each index $k$ across all $B$ subsequences using another BLSTM with recurrent batch normalization: 
	\begin{align}
	(\bm{\hat{o}}^{(1)}_k, \bm{\hat{o}}^{(2)}_k, \ldots, \bm{\hat{o}}^{(B)}_k) = BLSTM_{ws}(\bm{\bar{o}}^{(1)}_k, \bm{\bar{o}}^{(2)}_k, \ldots, \bm{\bar{o}}^{(B)}_k),
	\end{align}
	where $\bm{\hat{o}}^{(b)}_k \in \mathbb{R}^{H_{ws}}$. Similar to the intra-subsequence sequential modelling step, linear transformation via a fc layer, layer normalization, and a residual connection are then applied:
	\begin{align}
	\bm{{o}}^{(b)}_k = \bm{\hat{o}}^{(b)}_k + LN(\bm{W}_{ws}\bm{\hat{o}}^{(b)}_k + \bm{b}_{ws}), \label{eq:linear_transform_o}
	\end{align}
	resulting in the following $B$ output subsequences:
	\begin{align}
	\left( \begin{array}{cccc} \bm{{o}}^{(1)}_1 & \bm{{o}}^{(1)}_2 & \ldots & \bm{{o}}^{(1)}_K \\
	\bm{{o}}^{(2)}_1 & \bm{{o}}^{(2)}_2 & \ldots & \bm{{o}}^{(2)}_K \\
	\ldots & \ldots & \ldots & \ldots \\
	\bm{{o}}^{(B)}_1 & \bm{{o}}^{(B)}_2 & \ldots & \bm{{o}}^{(B)}_K \end{array} \right), \label{eq:output_subseq}
	\end{align}
	where $\bm{{o}}^{(b)}_k\!\in\!\mathbb{R}^{H_{ws}}$.  In (\ref{eq:linear_transform_o}), $\bm{W}_{ws}\!\!\in\!\mathbb{R}^{H_{ws}\times H_{ws}}$ and $\bm{b}_{ws}\!\in\!\mathbb{R}^{H_{ws}}$ denote the trainable weight matrix and bias vector of the fc layer, respectively. $\frac{H_{ws}}{2}$ is the size of the hidden states in $BLSTM_{ws}$ and we use the subscript $ws$ to indicate the modelling at the \emph{whole sequence} level. Given that an output vector $\bm{\bar{o}}^{(b)}_k$ contains the information of the entire $b$-th subsequence after the intra-subsequence sequential modelling step, an output vector $\bm{{o}}^{(b)}_k$ is expected to contain the information of all $B$ subsequences after the inter-subsequence sequential modelling step. In other words, $\bm{{o}}^{(b)}_k$ contains the information of the whole original long sequence of length $L$.
	
	Eventually, the output subsequences in (\ref{eq:output_subseq}) are unfolded, resulting in a single long sequence of $L$ elements $(\bm{{o}}_1, \bm{{o}}_2, \ldots, \bm{{o}}_L)$. After unfolding, the $k$-th element in the $b$-th subsequence, $\bm{{o}}^{(b)}_k$, in (\ref{eq:output_subseq}) will become the $\ell$-th element $\bm{{o}}_\ell$ in the final output sequence, where
	\begin{align}
	\ell = (b-1) \times K + k.
	\end{align}
	
	\subsection{Classification}
	
	For classification, the output vectors in the output sequence $(\bm{{o}}_1, \bm{{o}}_2, \ldots, \bm{{o}}_L)$ are passed through two fc layers, each with $N_{fc}$ units and Rectified Linear Unit (ReLU) activation, before being presented to an output layer with softmax activation to produce the network predictions $(\bm{\hat{y}}_1, \bm{\hat{y}}_2, \ldots, \bm{\hat{y}}_L)$. Note that these layers are shared across the time indices. 
	
	The network is trained to minimize the cross-entropy loss averaged over the sequence length and the training data:
	\begin{align} 
	\mathcal{L}(\bm{\theta}) = -\frac{1}{N\cdot L}\sum_{n=1}^N\sum_{\ell=1}^L \bm{y}^{(n)}_\ell\log\bm{\hat{y}}^{(n)}_\ell + \lambda||\bm{\theta}||^2_2. \label{eq:loss}
	\end{align}
	In (\ref{eq:loss}), $\bm{\theta}$ denotes the network parameters and $\lambda$ is the hyper-parameter of the $\ell_2$-norm regularization term.
	
	\section{Experiments}
	\label{sec:experiments}
	
	\subsection{Experimental setup}
	
	The experimental setup using the four databases described in Section \ref{sec:materials} is summarized in Table \ref{tab:exp_setup}. Conforming to majority of previous works in literature, we conducted leave-one-subject-out cross validation (LOSO CV) on SleepEDF, ear-EEG, and cEEGrid databases. For SHHS, we randomly split the subjects into 70\% for training and 30\% for testing. In each experiment, a number of subjects, specified in Table \ref{tab:exp_setup}, were held out from the training set for validation purpose. In particular, due to the small number of recordings in the SleepEDF and cEEGrid databases, we repeated these experiments 5 times and report the average performance. 
	
	\setlength\tabcolsep{2.25pt}
	\begin{table}[!b]
		\caption{Summary of experimental setup.}
		\begin{center}
			\begin{tabular}{|>{\arraybackslash}m{0.5in}|>{\centering\arraybackslash}m{0.4in}|>{\centering\arraybackslash}m{0.5in}|>{\centering\arraybackslash}m{0.8in}|>{\centering\arraybackslash}m{0.45in}|>{\centering\arraybackslash}m{0.45in}|>{\centering\arraybackslash}m{0in} @{}m{0pt}@{}}
				\cline{1-6}
				Database & \makecell{Num. of\\subjects} & \makecell{Num. of\\recordings} & Experimental setup & Held-out valid. set & Repetition & \parbox{0pt}{\rule{0pt}{1ex+\baselineskip}} \\ [0ex]  	
				\cline{1-6}
				SHHS & 5,463 & 5,463 & train/test: 0.7/0.3 & 100 & 1 & \parbox{0pt}{\rule{0pt}{0ex+\baselineskip}} \\ [0ex]  	
				SleepEDF & 20 & 39 & LOSO CV & 4 & 5 &\parbox{0pt}{\rule{0pt}{0ex+\baselineskip}} \\ [0ex]  	
				ear-EEG & 20 & 75  & LOSO CV & 4 & 1 &\parbox{0pt}{\rule{0pt}{0ex+\baselineskip}} \\ [0ex]  	
				cEEGrid & 12 & 12 & LOSO CV & 2 & 5 &\parbox{0pt}{\rule{0pt}{0ex+\baselineskip}} \\ [0ex]  	
				\cline{1-6}
			\end{tabular}
		\end{center}
		\label{tab:exp_setup}
	\end{table}
	
	\subsection{Parameters}
	
	We experimented with the sequence length $L=200$ epochs (equivalent to 100 minutes of sleep data to roughly cover a whole sleep cycle), the number of subsequences $B=10$, and  the subsequence length $K=20$. We also experimented with other values for $L$, $B$, and $K$, and will discuss their influence in Section \ref{sssec:influence_seq_len}. Regarding the network architecture, we set the number of filters $M=32$, the attention size $A=64$, the size of BLSTM's hidden states $\frac{H_e}{2} = \frac{H_{ss}}{2} = \frac{H_{ws}}{2} = 64$, and the size of the fc layers $N_{fc} = 512$. The hyper-parameter $\lambda$ in (\ref{eq:loss}) was fixed to $10^{-4}$. In addition, a dropout rate of 0.1 was applied to the LSTM cells and the fc layers during training.
	
	The network was trained using Adam optimizer \cite{Kingma2015} with a learning rate of $10^{-4}$, $\beta_1 = 0.9$, $\beta_2 = 0.999$, and $\epsilon= 10^{-7}$. A minibatch size of 8 sequences was used for training. Similar to prior works (e.g., SeqSleepNet \cite{Phan2019a}, XSleepNet \cite{Phan2022a}, and SleepTransformer \cite{Phan2022c}), the sequences were sampled with a shift of one epoch. The model was validated on the validation set every 100 training steps for the experiments on SHHS, SleepEDF, and ear-EEG. For the smallest database cEEGrid, the validation was done every 25 training steps. For the largest database SHHS, the model was trained for fixed 5000 validation steps without early stopping. Other than that, the model was trained for 10 training epochs and stopped early after 50 validation steps (for SleepEDF and ear-EEG) and 25 validation steps (for cEEGrid) without improvement on the validation accuracy. The model performing best on the validation set was then retained to be evaluated on the test data. 
	
	\subsection{The baseline}
	
	We used SeqSleepNet presented in \cite{Phan2019a} as the main baseline in the experiments. This network shares a similar input format and epoch encoding component as the proposed L-SeqSleepNet. However, different from L-SeqSleepNet, it follows the typical ``flat'' sequential modelling approach as many other works in literature, making it a natural baseline to investigate the effects of the long sequential modelling approach presented here. We compared L-SeqSleepNet and the SeqSleepNet baseline under two initialization schemes: (1) random initialization (i.e., training from scratch) and (2) initialization with a pretrained network (i.e. finetuning for transfer learning \cite{Phan2020b}). For the former, the networks were trained from scratch as usual. For the latter, the networks trained on SHHS were used as the pretrained models and finetuned on the SleepEDF, ear-EEG, and cEEGrid databases. 
	
	In addition, in order to assess L-SeqSleepNet in terms of sleep staging performance, we also compare it to other methods in literature reporting results on the experimental databases. 
	
	\setlength\tabcolsep{2pt}
	\begin{table*}[!t]
		\caption{Performance obtained by L-SeqSleepNet ($L=200$), the SeqSleepNet baseline ($L=20$) in comparison with previous works on the experimental databases. Note that, on SHHS, the results with L-SeqSleepNet$^\diamond$ are without data exclusion (see Section \ref{sec:materials}). Furthermore, some results reported in previous works, marked by the $^\ddagger$ superscript,  are not compatible for a direct comparison here due to the discrepancies in data split, the number of channel used, the number of subject used, modelling tasks, etc. The $^*$ superscript indicates the pretraining-based initialization scheme.}
		\scriptsize
		\begin{center}
			\begin{tabular}{|>{\centering\arraybackslash}m{0.4in}|>{\arraybackslash}m{0.95in}|>{\centering\arraybackslash}m{0.475in}|>{\centering\arraybackslash}m{0.675in}|>{\centering\arraybackslash}m{0.475in}|>{\centering\arraybackslash}m{0.475in}|>{\centering\arraybackslash}m{0.475in}||>{\centering\arraybackslash}m{0.475in}|>{\centering\arraybackslash}m{0.475in}|>{\centering\arraybackslash}m{0.475in}|>{\centering\arraybackslash}m{0.475in}|>{\centering\arraybackslash}m{0.475in}|>{\centering\arraybackslash}m{0in} @{}m{0pt}@{}}
				\cline{1-12}
				\multirow{2}{*}{\makecell{Database}} &  \multirow{2}{*}{\makecell{System}} &  \multicolumn{5}{c||}{Overall performance} & \multicolumn{5}{c|}{Class-wise MF1} & \parbox{0pt}{\rule{0pt}{2ex+\baselineskip}} \\ [0ex]  	
				\cline{3-12}
				& & Acc. & $\kappa$ & MF1 & Sens. & Spec. & Wake & N1 & N2 & N3 & REM & \parbox{0pt}{\rule{0pt}{2ex+\baselineskip}} \\ [0ex]  	
				
				\cline{1-12}
				\multirow{11}{*}{\makecell{SHHS}} & \bf L-SeqSleepNet  &  ${88.4}$ & ${0.838}$ & ${81.4}$ & ${80.4}$ & ${96.7}$ &  ${93.1}$ & $51.1$ & ${89.0}$ & $84.9$ & ${89.8}$ & \parbox{0pt}{\rule{0pt}{0ex+\baselineskip}} \\ [0ex]  	
				& \bf L-SeqSleepNet$^\diamond$  &  $\it {87.6}$ & $\it {0.825}$ & $\it {80.3}$ & $\it {79.4}$ & $\it {96.5}$ &  $\it {92.4}$ & $\it 48.6$ & ${\it 88.2}$ & $\it 83.9$ & $\it {88.5}$ & \parbox{0pt}{\rule{0pt}{0ex+\baselineskip}} \\ [0ex]  	
				
				& \bf \emph{SeqSleepNet \cite{Phan2019a}} & $87.2$ & $0.820$ & $80.2$ & $78.7$ & $96.3$ & $91.8$ & $49.1$ & $88.2$ & $83.5$ & $88.2$ & \parbox{0pt}{\rule{0pt}{0ex+\baselineskip}} \\ [0ex]  	
				
				& SleepTransformer \cite{Phan2022c} & ${87.7}$ & ${0.828}$ & $80.1$ & $78.7$ & ${96.5}$ & $92.2$ & $46.1$ & $88.3$ & ${85.2}$ & ${88.6}$&\parbox{0pt}{\rule{0pt}{0ex+\baselineskip}} \\ [0ex]  	
				
				& XSleepNet1 \cite{Phan2022a}  & {${87.6}$} & {${0.826}$} & {${80.7}$} & {$79.7$} & {${96.5}$} & ${91.6}$ & ${51.4}$ & ${88.5}$ & ${85.0}$ & ${88.4}$ & \parbox{0pt}{\rule{0pt}{0ex+\baselineskip}} \\ [0ex]  	
				
				& XSleepNet2 \cite{Phan2022a}  & ${87.5}$ & ${0.826}$ & ${81.0}$ & ${80.4}$ & ${96.5}$  &  {${92.0}$} & {${49.9}$} & {${88.3}$} & {${85.0}$} & {${88.2}$} &\parbox{0pt}{\rule{0pt}{0ex+\baselineskip}} \\ [0ex]  	
				
				& U-Sleep$^\ddagger$ \cite{Perslev2021} & $-$ & $-$ & $80.0$ & $-$ & $-$ & $93.0$ & $51.0$ & $87.0$ & $76.0$ & $92.0$ &  \parbox{0pt}{\rule{0pt}{0.25ex+\baselineskip}} \\ [0ex]  	
				
				& Olesen \emph{et al.}$^\ddagger$ \cite{Olesen2021} & $87.1$ & $0.816$ & $78.8$ & $77.7$ & $96.3$ & $94.1$ & $47.8$ & $87.9$ & $74.3$ & $89.9$ &  \parbox{0pt}{\rule{0pt}{0.25ex+\baselineskip}} \\ [0ex]  	
				
				& CNN \cite{Sors2018} & $86.8$ & $0.810$ & $78.5$ & $-$ & $95.0$ & $-$ & $-$ & $-$ & $-$ & $-$ &  \parbox{0pt}{\rule{0pt}{0.25ex+\baselineskip}} \\ [0ex]  	
				
				& FCNN+RNN \cite{Phan2022a}  &   $ 86.7$ & $ 0.813$ & $ 79.5$ & $ 78.1$ & $ 96.2$ & $ 91.1$ & $ 48.7$ & $ 88.0$ & $ 82.6$ & $ 87.1$ &  \parbox{0pt}{\rule{0pt}{0.25ex+\baselineskip}} \\ [0ex]  	
				
				&  IITNet \cite{Seo2020} & $86.7$ & $0.810$ & $79.8$ & $-$ & $-$ & $-$ & $-$ & $-$ & $-$ & $-$ &  \parbox{0pt}{\rule{0pt}{0.25ex+\baselineskip}} \\ [0ex]  	
				
				& AttnSleep$^\ddagger$ \cite{Eldele2021}  & $84.2 $ & $0.780$ & $75.3$ & $-$ & $-$ & $86.7$ & $33.2$ & $87.1$ & $87.1$ & $82.1$ &  \parbox{0pt}{\rule{0pt}{0.25ex+\baselineskip}} \\ [0ex]  	
				
				\cline{1-12}
				
				\multirow{23}{*}{\makecell{SleepEDF}} & \bf L-SeqSleepNet$^*$  & $88.6\pm0.1$ & $0.845\pm0.001$ & $82.9\pm0.2$ & $82.1\pm0.1$ & $96.9\pm0.0$ & $94.1\pm0.4$ & $53.3\pm1.2$ & $89.7\pm0.1$ & $88.4\pm0.3$ & $88.9\pm0.2$ & \parbox{0pt}{\rule{0pt}{0ex+\baselineskip}} \\ [0ex]  	
				
				&\bf L-SeqSleepNet & $86.3\pm0.2$ & $0.813\pm0.003$ & $79.3\pm0.4$ & $78.8\pm0.5$ & $96.3\pm0.1$  & $91.6\pm0.4$ & $45.3\pm1.4$ & $88.5\pm0.3$ & $86.2\pm0.7$ & $85.2\pm0.2$ & \parbox{0pt}{\rule{0pt}{0ex+\baselineskip}} \\ [0ex]  	
				
				&\bf \emph{SeqSleepNet$^*$ \cite{Phan2019a}}  &  $87.6\pm0.2$ & $0.830\pm0.002$ & $81.8\pm0.2$ & $80.3\pm0.3$ & $96.6\pm0.1$  & $92.7\pm0.4$ & $52.7\pm0.7$ &  $88.9\pm0.1$ & $86.7\pm0.2$ & $87.8\pm0.1$ & \parbox{0pt}{\rule{0pt}{0ex+\baselineskip}} \\ [0ex]  	
				
				&\bf \emph{SeqSleepNet \cite{Phan2019a}}  &$85.6\pm0.3$ &$0.803\pm 0.004$ &$78.6\pm 0.2$ &$78.2\pm 0.1$ &$96.2\pm 0.1$ &$91.2\pm 0.6$ &$44.7\pm 0.8$ &$88.0\pm 0.1$ &$86.2\pm 0.2$ &$83.0\pm 0.8$ & \parbox{0pt}{\rule{0pt}{0ex+\baselineskip}} \\ [0ex]  	
				
				& SalientSleepNet$^\ddagger$ \cite{Jia2021}  & $87.5$ & $-$ & $83.0$ & $-$ & $-$  & $92.3$ & $56.2$ & $89.9$ & $87.2$ & $89.2$ & \parbox{0pt}{\rule{0pt}{0ex+\baselineskip}} \\ [0ex]  	
				
				& TransSleep \cite{Phyo2022} & $86.5$ & $-$ & $82.5$ & $-$ & $-$  & $87.1$ & $60.8$ & $91.7$ & $85.5$ & $87.4$ & \parbox{0pt}{\rule{0pt}{0ex+\baselineskip}} \\ [0ex]  	
				
				& XSleepNet2 \cite{Phan2022a} &  ${86.3}$ & ${0.813}$ & ${80.6}$ & ${80.2}$ & ${96.4}$  &  ${92.2}$ & ${51.8}$ & $88.0$ & ${86.8}$ & ${83.9}$ & \parbox{0pt}{\rule{0pt}{0ex+\baselineskip}} \\ [0ex]  	
				
				& XSleepNet1 \cite{Phan2022a}  &  ${86.0}$ & ${0.810}$ & ${80.0}$ & ${79.6}$ & ${96.3}$   &  $91.3$ & ${49.5}$ & $88.0$ & ${86.9}$ & ${84.2}$ & \parbox{0pt}{\rule{0pt}{0ex+\baselineskip}} \\ [0ex]  	
				
				& SimpleSleepNet$^\ddagger$ \cite{Guillot2020} & $-$ & $-$ & $80.5$ & $-$ & $-$  & $-$ & $-$ & $-$ & $-$ & $-$ & \parbox{0pt}{\rule{0pt}{0ex+\baselineskip}} \\ [0ex]  	

				& MNN$^\ddagger$ \cite{Dong2017} & $85.9$ & $-$ & $80.5$ & $-$ & $-$  & $84.6$ & $56.3$ & $90.7$ & $84.8$ & $86.1$ & \parbox{0pt}{\rule{0pt}{0ex+\baselineskip}} \\ [0ex]  	
				
				& Khalili \& Asl \cite{Khalili2021}  & $85.4$ & $0.800$ & $79.3$ & $-$ & $-$  & $90.0$ & $46.6$ & $88.4$ & $86.1$ & $84.6$ & \parbox{0pt}{\rule{0pt}{0ex+\baselineskip}} \\ [0ex]  	
				
				& TinySleepNet \cite{Supratak2020} & $85.4$ & $0.800$ & $80.5$ & $-$ & $-$  & $90.1$ & $51.4$ & $88.5$ & $88.3$ & $84.3$ & \parbox{0pt}{\rule{0pt}{0ex+\baselineskip}} \\ [0ex]  	
				
				& RobustSleepNet$^\ddagger$ \cite{Guillot2021}  & $-$ & $-$ & $79.1$ & $-$ & $-$  & $-$ & $-$ & $-$ & $-$ & $-$ & \parbox{0pt}{\rule{0pt}{0ex+\baselineskip}} \\ [0ex]  	
				
				& U-Sleep \cite{Perslev2021}  & $-$ & $-$ & $79.0$ & $-$ & $-$ & $93.0$ & $57.0$ & $86.0$ & $71.0$ & $88.0$ & \parbox{0pt}{\rule{0pt}{0ex+\baselineskip}} \\ [0ex]  	
				
				& SleepFCN \cite{Goshtasbi2022}  & $84.8$ & $0.780$ & $78.8$ & $-$ & $-$  & $89.6$ & $44.6$ & $89.1$ & $90.6$ & $80.3$ & \parbox{0pt}{\rule{0pt}{0ex+\baselineskip}} \\ [0ex]  	
				
				& MRASleepNet \cite{Yu2022}  & $84.5$ & $0.786$ & $78.9$ & $-$ & $-$  & $-$ & $-$ & $-$ & $-$ & $-$ & \parbox{0pt}{\rule{0pt}{0ex+\baselineskip}} \\ [0ex]  	
				
				& ResNetMHA \cite{Qu2020}  & $84.3$ & $-$ & $79.0$ & $-$ & $-$  & $90.2$ & $48.3$ & $87.8$ & $85.6$ & $83.3$ & \parbox{0pt}{\rule{0pt}{0ex+\baselineskip}} \\ [0ex]  	
				
				& AttnSleep \cite{Eldele2021}  & $84.4$ & $0.790$ & $78.1$ & $-$ & $-$  & $89.7$ & $42.6$ & $88.8$ & $90.2$ & $79.0$ & \parbox{0pt}{\rule{0pt}{0ex+\baselineskip}} \\ [0ex]  	
				
				& DeepSleepNet-Lite \cite{Fiorillo2021}  & $84.0$ & $0.780$ & $78.0$ & $-$ & $-$ & $87.1$ & $44.4$ & $87.9$ & $88.2$ & $82.4$ &  \parbox{0pt}{\rule{0pt}{0.25ex+\baselineskip}} \\ [0ex]  	

				& IITNet \cite{Seo2020} & $83.9$ & $0.780$ & $77.6$ & $-$ & $-$  & $-$ & $-$ & $-$ & $-$ & $-$ & \parbox{0pt}{\rule{0pt}{0ex+\baselineskip}} \\ [0ex]  	

				& DeepSleepNet \cite{Supratak2017}  & $82.0$ & $0.760$ & $76.9$  & $-$ & $-$   & $86.7$ & $45.5$ & $85.1$ & $83.3$ & $82.6$ & \parbox{0pt}{\rule{0pt}{0ex+\baselineskip}} \\ [0ex]  	
				
				& FCNN+RNN \cite{Phan2022a}  & $81.8$ & $0.754$ & $ 75.6$ & $75.7$ & $95.3$   &  $ 89.4$ & $ 44.1$ & $ 84.0$ & $ 84.0$ & $ 76.3$ & \parbox{0pt}{\rule{0pt}{0ex+\baselineskip}} \\ [0ex]  	
				
				& SleepEEGNet \cite{MousaviI2019} & $81.5$ & $0.750$ & $76.6$ & $-$ & $-$  & $89.4$ & $44.4$ & $84.7$ & $84.6$ & $79.6$ & \parbox{0pt}{\rule{0pt}{0ex+\baselineskip}} \\ [0ex]  	

				\cline{1-12}
				
				\multirow{6}{*}{\makecell{earEEG}} & \bf L-SeqSleepNet*  & $87.9$ & $0.829$ & $84.1$ & $83.1$ & $96.5$  & $92.7$ & $59.2$ & $89.8$ & $89.4$ & $89.2$ & \parbox{0pt}{\rule{0pt}{0ex+\baselineskip}} \\ [0ex]  	
				
				& \bf L-SeqSleepNet  & $83.7$ & $0.770$ & $79.4$ & $78.7$ & $95.3$  &  $89.3$ & $52.3$ & $86.1$ & $87.4$ & $81.9$ &\parbox{0pt}{\rule{0pt}{0ex+\baselineskip}} \\ [0ex]  	
				
				& \bf \emph{SeqSleepNet* \cite{Phan2019a}} &  $86.0$ & $0.801$ & $81.9$ & $80.3$ & $95.8$  & $90.7$ & $55.7$ & $88.2$ & $88.3$ & $86.5$ & \parbox{0pt}{\rule{0pt}{0ex+\baselineskip}} \\ [0ex]  	
				
				& \bf \emph{SeqSleepNet \cite{Phan2019a}}  & $83.0$   & $0.759$ & $78.5$ & $77.3$ & $95.0$  & $89.0$ & $50.0$ & $85.7$ & $87.9$ & $79.8$ & \parbox{0pt}{\rule{0pt}{0ex+\baselineskip}} \\ [0ex]  	
				
				& Ensemble \cite{Borup2023}  & $-$   & $0.780$ & $-$ & $-$ & $-$  & $-$ & $-$ & $-$ & $-$ & $-$ & \parbox{0pt}{\rule{0pt}{0ex+\baselineskip}} \\ [0ex]  	
				
				& Random Forest$^\ddagger$ \cite{Mikkelsen2019b} & $-$   & $0.730$ & $-$ & $-$ & $-$  & $-$ & $-$ & $-$ & $-$ & $-$ & \parbox{0pt}{\rule{0pt}{0ex+\baselineskip}} \\ [0ex]  	
				
				\cline{1-12}
				
				\multirow{9}{*}{\makecell{cEEGrid}} & \bf L-SeqSleepNet* & $78.9\pm0.6$ & $0.703\pm0.008$ & $67.9\pm0.5$ & $67.0\pm0.3$ & $94.2\pm0.1$  & $92.0\pm0.7$ & $25.8\pm1.4$ & $74.1\pm0.6$ & $79.7\pm1.1$ & $68.0\pm1.5$ &\parbox{0pt}{\rule{0pt}{0ex+\baselineskip}} \\ [0ex]  	
				
				& \bf L-SeqSleepNet  & $72.3\pm0.6$ & $0.607\pm0.008$ & $57.3\pm0.8$ & $57.8\pm0.9$ & $92.4\pm0.2$  & $89.9\pm0.6$ & $7.5\pm1.8$ & $65.8\pm0.9$ & $72.4\pm1.8$ & $50.8\pm2.4$ & \parbox{0pt}{\rule{0pt}{0ex+\baselineskip}} \\ [0ex]  	
				
				& \bf \emph{SeqSleepNet* \cite{Phan2019a}}  & $75.0\pm0.4$  & $0.647\pm0.006$ & $62.7\pm0.7$ & $61.3\pm0.8$ & $93.2\pm0.1$  & $90.6\pm0.5$ & $23.1\pm1.1$ & $71.1\pm0.5$ & $72.1\pm0.7$ & $56.6\pm3.1$ & \parbox{0pt}{\rule{0pt}{0ex+\baselineskip}} \\ [0ex]  	
				
				& \bf \emph{SeqSleepNet \cite{Phan2019a}}  &  $70.4\pm1.5$ & $0.578\pm0.024$ & $54.1\pm1.5$ & $55.0\pm1.7$ & $91.7\pm0.5$  & $87.4\pm1.9$ & $0.5\pm0.5$ & $64.6\pm1.7$ & $69.3\pm1.7$ & $48.6\pm5.0$ & \parbox{0pt}{\rule{0pt}{0ex+\baselineskip}} \\ [0ex]  	
				
				& ADA pers$^\ddagger$ \cite{Heremans2022a}  &  $72.8$ & $0.618$ & $-$ & $-$ & $-$  & $-$ & $-$ & $-$ & $-$ & $-$ &  \parbox{0pt}{\rule{0pt}{0ex+\baselineskip}} \\ [0ex]  	
				
				& Feat. matching$^\ddagger$ \cite{Heremans2022b} &  $71.3$ & $0.605$ & $-$ & $-$ & $-$  & $-$ & $-$ & $-$ & $-$ & $-$ &  \parbox{0pt}{\rule{0pt}{0ex+\baselineskip}} \\ [0ex]  	
				
				& Random Forest$^\ddagger$ \cite{Mikkelsen2019}  &  $70.0$ & $0.580$ & $-$ & $-$ & $-$  & $-$  & $-$ & $-$ & $-$ & $-$ &  \parbox{0pt}{\rule{0pt}{0ex+\baselineskip}} \\ [0ex]  	
				
				& DeepSleepNet* \cite{Phan2020b} &  $58.2$ & $0.391$ & $42.8$ & $-$ & $-$  & $74.9$ & $5.7$ & $47.0$ & $63.4$ & $23.3$ &  \parbox{0pt}{\rule{0pt}{0ex+\baselineskip}} \\ [0ex]  	
				
				& DeepSleepNet \cite{Phan2020b} &  $42.5$ & $0.195$ & $30.3$ & $-$ & $-$  & $57.6$  & $6.9$ & $23.1$ & $51.4$ & $12.3$ &  \parbox{0pt}{\rule{0pt}{0ex+\baselineskip}} \\ [0ex]  	
				\cline{1-12}
			\end{tabular}
		\end{center}
		\label{tab:performance}
	\vspace{-0.2cm}
	\end{table*}
	
	\begin{figure*} [!t]
		\centering
		\includegraphics[width=0.95\linewidth]{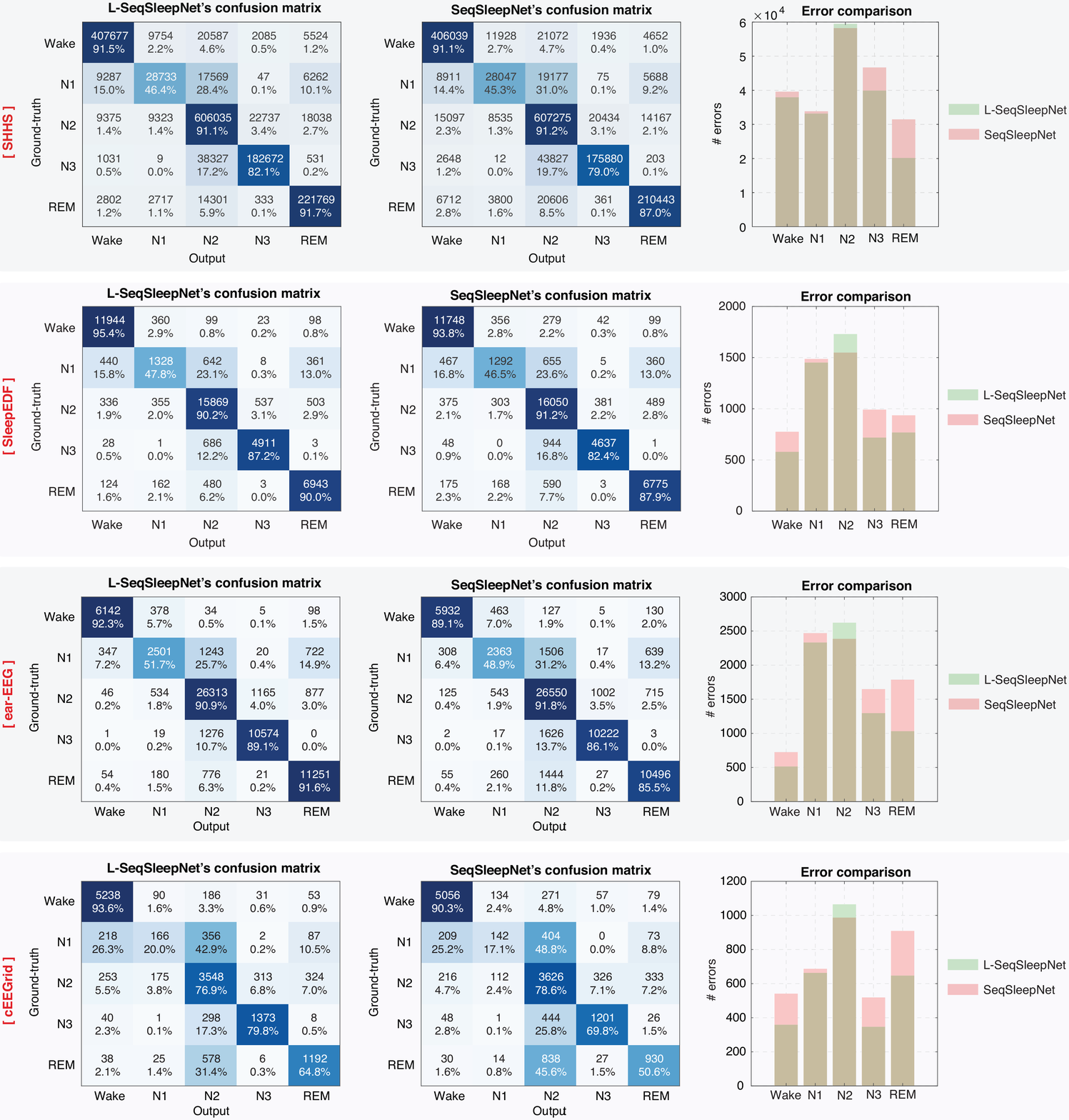}
		\caption{Comparison between L-SeqSleepNet ($L=200$) and SeqSleepNet ($L=20$).}
		\label{fig:confusionmatrix}
		\vspace{-0.2cm}
	\end{figure*}
	
	\subsection{Experimental results}
	\label{sec:experiment_results}
	
	\subsubsection{Overall sleep staging performance}
	A comprehensive performance comparison between L-SeqSleepNet, the SeqSleepNet baseline, and prior works are shown in Table \ref{tab:performance}. On the one hand, it can be seen that L-SeqSleepNet consistently outperforms the SeqSleepNet baseline over all the experimental databases. With the random initialization scheme, L-SeqSleepNet leads to overall accuracy gains of $1.2\%$, $0.7\%$, $0.7\%$, and $1.9\%$ on SHHS, SleepEDF, ear-EEG, and cEEGrid, respectively. In case of the pretraining-based initialization, the gains on SleepEDF, ear-EEG, and cEEGrid are even higher, reaching $1.0\%$, $1.9\%$, and $3.9\%$, respectively. The wider gains obtained with respect to the pretraining-based initialization shed some light on what is being transferred from the source domain (i.e. SHHS) to the target domains (i.e., SleepEDF, ear-EEG, and cEEGrid) via L-SeqSleepNet in these transfer learning scenarios. More specifically, in addition to the usual reuse of good feature representations \cite{Neyshabur2020}, it is likely that the diverse sleep cycle structure from the large cohort in the source domain also contributes to the transferred knowledge and gives rise to L-SeqSleepNet's higher performance gains compared to the SeqSleepNet baseline. 
	
	On the other hand, L-SeqSleepNet also results in better performance than the current state-of-the-arts (where the direct comparison is compatible) over all the databases. On the large database SHHS, L-SeqSleepNet achieves an overall accuracy of $88.4\%$, $0.7\%$ and $0.8\%$ higher than SleepTransformer \cite{Phan2022c} and XSleepNet \cite{Phan2022a}, respectively. This is particularly interesting given that the SeqSleepNet-type architecture of L-SeqSleepNet essentially constitutes only one half of the XSleepNet's architecture \cite{Phan2022a} and that L-SeqSleepNet has around $6.3\!\times\!10^5$ parameters in total, roughly 9 times smaller than XSleepNet which has $5.7\!\times\!10^6$ parameters. On the smaller databases (i.e., SleepEDF, ear-EEG, and cEEGrid), a large margin of performance is consistently seen between L-SeqSleepNet and other counterparts. 
	
	\subsubsection{The effects of whole-cycle sequence modelling}
	
	Using the pretraining-based initialization scheme, we inspected the effects of L-SeqSleepNet's whole-cycle sequence modelling to the model's errors.  From the confusion matrices and the number of errors per sleep stage in Figure \ref{fig:confusionmatrix}, it becomes clear that across all the databases, compared to the SeqSleepNet baseline, L-SeqSleepNet lowers the errors in all other sleep stages, most noticeably in N3 and REM, at the small expense of errors in N2. This implies that taking into account the structure of sleep cycles helps to correct modelling errors that are impossible to fix under the existing state-of-the-art (short) context modelling approach. 
	
\begin{figure} [!t]
		\centering
		\includegraphics[width=1\linewidth]{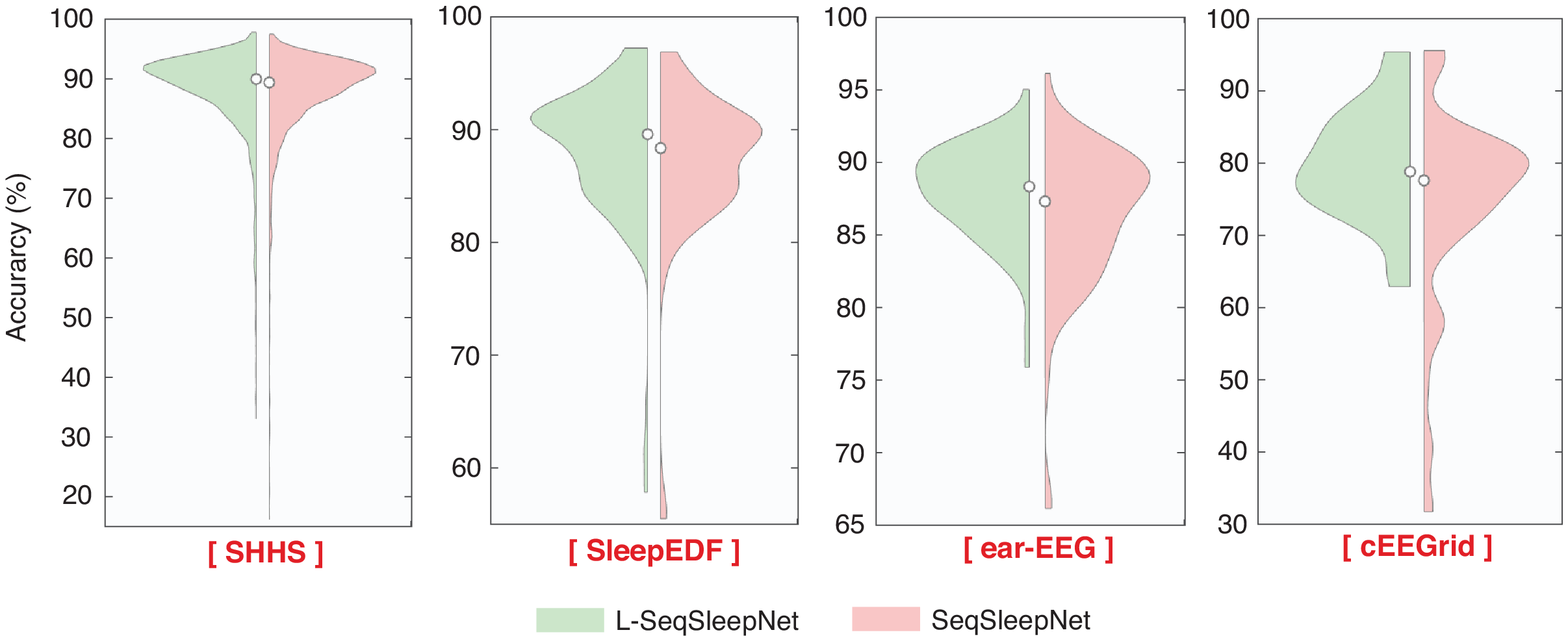} 
		\caption{Distribution of individual accuracies produced by L-SeqSleepNet and the SeqSleepNet baseline, demonstrating the robustness of L-SeqSleepNet. }
		\label{fig:individual_errors}
	\end{figure}

	Figure \ref{fig:individual_errors} visually represents the distribution of individual errors produced by L-SeqSleepNet and the SeqSleepNet baseline. It was found in \cite{Baumert2023} that even though the performance of automatic sleep staging systems has been deemed sufficient for clinical use these systems may generally struggle with particular recordings, leading to exceptionally high individual errors. The distribution of individual errors from both L-SeqSleepNet and the baseline in the figure indeed reflects this finding, particularly on SHHS. However, L-SeqSleepNet can reduce these exceptionally high individual errors across all the databases, and more strikingly on ear-EEG and cEEGrid. This is particularly important from an application point of view as bringing down these high individual errors would make automatic sleep staging systems more acceptable in clinical environments as well as require less human intervention for manual editing or rescoring. 
	
	\subsubsection{The influence of the sequence length}
	\label{sssec:influence_seq_len}
	In this section, we examined the influence of the sequence length to L-SeqSleepNet and the SeqSleepNet baseline from two perspectives: staging performance and computational time. Using the large database SHHS in this examination, Table \ref{tab:effect_seq_len} summarizes the overall sleep staging performance and computational time of the two models with different sequence lengths. 
	
	Firstly, as already mentioned in Section \ref{sec:introduction}, SeqSleepNet with the ``flat'' sequential modelling approach is inefficient in handling long sequences, resulting in marginal gains on overall accuracy, for instance, $0.3\%$ when the sequence length increases to 200 from 20 epochs. In contrast, at the sequence length of 200 epochs, the improvement on overall accuracy achieved by L-SeqSleepNet over SeqSleepNet with $L=20$ reaches $1.2\%$, highlighting the effectiveness of the proposed long sequential modelling approach. 
	
	Secondly, concerning L-SeqSleepNet itself, halving the sequence length to 100 epochs (50 minutes, roughly a half sleep cycle) results in a drop of $0.3\%$ on overall accuracy while doubling it to 400 epochs (200 minutes, roughly two sleep cycles) causes negligible consequence to the performance. This suggests that the sequential information in one sleep cycle is probably all we need for sleep staging.
	Interestingly, the way a long sequence is folded into subsequences seems to matter. At $L=200$, folding with $B=20$ and $K=10$ (i.e., 20 subsequences, each of length 10 epochs) leads to better performance than that with $B=10$ and $K=20$ (i.e., 10 subsequences, each of length 20 epochs). It is likely the distance between two adjacent elements in the sequence during inter-subsequence sequential modelling causes the difference. In effect, this distance is 10 epochs in the former while it is 20 epochs in the latter. The double distance in the latter could loosen the temporal dependency between the elements in the sequence, worsening the performance consequentially.  
	
	Thirdly, given that the SeqSleepNet baseline's training time grows linearly as the sequence length increases, L-SeqSleepNet's training time only grows sub-linearly\footnote{The benchmark was run on a NVIDIA DGX-1 machine using a single NVIDIA Tesla V100 GPU.}. For example, SeqSleepNet at $L=200$ requires 1,465 seconds for 1000 training steps, 4.5 times longer than itself at $L=20$. L-SeqSleepNet at $L=200$, on the other hand, merely needs around 450 seconds, just 1.4 times slower than SeqSleepNet at $L=20$ and around 3.3 times faster than SeqSleepNet at $L=200$. The reason is when a sequence of length $L$ is folded into $B$ subsequences, each of length $K$, the number of time steps engaged in sequential modelling is in fact reduced to $B+K$ which is much smaller than $L$.

	\section{Discussion}
	\label{sec:discussion}
	This work demonstrates the importance and potential of long sequence modelling. This is natural as sleep cycles last for around 90 minutes, and we have demonstrated that taking this into account improves automatic sleep staging accuracy and robustness. In particular, we also see that the kappa value $0.829$ obtained by L-SeqSleepNet is \textit{higher} than the human inter-scorer agreement \cite{Rosenberg2013}. This comparison is based on the ear-EEG database, where manual scoring from two independent sleep technicians are available, similar to what was presented in Mikkelsen \emph{et al.} \cite{Mikkelsen2021}. This result is appealing given the fact that L-SeqSleepNet only used the bilateral ear-EEG derivation, and thus, did not have access to any of the same electrode derivations as were used in the manual scoring of this database. This indicates that long time scale sleep information can compensate for a decrease in ``field of view'' on the sensor level, which is remarkable. This observation also suggests that consolidating intrinsic information from a long context, like a whole sleep cycle or more, is something an automated scorer can excel at, resembling how clinicians use the knowledge about the sleep cycles and additional information available in form of signal trends to keep track of where in a sleep cycle they are in during manual scoring.
	
	
	While the work here focuses on single-EEG input and uses RNN as the backbone for sequential modelling, the presented long sequential modelling method can be considered as a generic method and can be used in replacement for the ``flat'' sequential modelling method in the current state-of-the-art sequence-to-sequence sleep staging framework \cite{Phan2022b}. It can be readily integrated into existing works relying on this framework to investigate the effects of whole-cycle sequential modelling in multimodal fusion \cite{Guillot2021} and multi-view learning \cite{Phan2022a} or to speed up the computation with recurrent-free architectures, such as Transformer \cite{Phan2022c}.

	\setlength\tabcolsep{2.25pt}
	\begin{table}[!t]
		\caption{The overall performance and the training time produced by L-SeqSleepNet and the SeqSleepNet baseline with different sequence lengths. Note that the training time was measured for 1000 training steps.}
		\begin{center}
			\begin{tabular}{|>{\arraybackslash}m{0.7in}|>{\centering\arraybackslash}m{0.65in}|>{\centering\arraybackslash}m{0.3in}|>{\centering\arraybackslash}m{0.3in}|>{\centering\arraybackslash}m{0.3in}|>{\centering\arraybackslash}m{0.45in}|>{\centering\arraybackslash}m{0in} @{}m{0pt}@{}}
				\cline{1-6}
				System & $L~(B \times K)$ & Acc. & $k$ & MF1 & \makecell{Training\\time (s)} &\parbox{0pt}{\rule{0pt}{1ex+\baselineskip}} \\ [0ex]  	
				\cline{1-6}
				L-SeqSleepNet & $100~(10\times10)$ &  $88.1$ & $0.833$ & $80.7$  & 354.1 & \parbox{0pt}{\rule{0pt}{0.5ex+\baselineskip}} \\ [0ex]  	
				L-SeqSleepNet & $200~(10\times20)$ & $88.2$ & $0.835$ & $81.4$ & 454.4 & \parbox{0pt}{\rule{0pt}{0.5ex+\baselineskip}} \\ [0ex]  	
				L-SeqSleepNet & $200~(20\times10)$ &  $88.4$ & $0.837$ & $81.6$ & 449.0 & \parbox{0pt}{\rule{0pt}{0.5ex+\baselineskip}} \\ [0ex]  	
				L-SeqSleepNet & $400~(20\times20)$ & $88.4$ & $0.838$ & $81.4$  & 605.2 & \parbox{0pt}{\rule{0pt}{0.5ex+\baselineskip}} \\ [0ex]  	
				\cline{1-6}
				SeqSleepNet & 20 &  $87.2$ & $0.820$ & $80.2$ & 322.1 & \parbox{0pt}{\rule{0pt}{0.5ex+\baselineskip}} \\ [0ex]  	
				SeqSleepNet & 100 & $87.4$ & $0.823$ & $80.1$ & 842.0 & \parbox{0pt}{\rule{0pt}{0.5ex+\baselineskip}} \\ [0ex]  	
				SeqSleepNet & 200 & $87.5$ & $0.824$ & $80.4$ & 1465.6 & \parbox{0pt}{\rule{0pt}{0.5ex+\baselineskip}} \\ [0ex]  	
				\cline{1-6}
			\end{tabular}
		\end{center}
		\label{tab:effect_seq_len}
	\end{table}

	\section{Conclusions}
	
	We presented in this work a novel method for modelling whole-cycle long sequences for automatic sleep staging. A long sequence was first folded into multiple subsequences. Intra-subsequence and inter-subsequence sequential modelling were then performed before the subsequences were unfolded to resume the size of the original sequence. We demonstrated that the proposed approach overcomes the limitations of the existing sequential modelling approach in handling long sequences and that taking the structural information of sleep cycles into account consistently improved the sleep staging performance. L-SeqSleepNet, the network with long sequential modelling capacity we introduced, outperformed not only the baseline but also the existing state-of-the-art methods across four distinct databases with different EEG setups, including scalp EEG, in-ear EEG, and around-the-ear EEG. We also empirically showed that incorporating the logic of stage transition in sleep cycles helped to reduce staging errors at epoch level as well as improves overall accuracy in the most difficult recordings (as reflected by exceptionally high errors of the baseline model). Furthermore, the performance benefits came at just a little cost of sub-linear growth in computational overhead.
	
	\section*{Acknowledgment}
	This research received funding from the Flemish Government (AI Research Program FLAIR) and from FWO Research Project G0D8321N . Maarten De Vos is affiliated to Leuven.AI - KU Leuven institute for AI, B-3000, Leuven, Belgium. H. Phan was supported by a Turing Fellowship under the EPSRC grant EP/N510129/1.
	\vspace{-0.2cm}
	\bibliographystyle{IEEEbib}
	\bibliography{bibliography}
	
\end{document}